\documentclass{article}
\usepackage{ismir,amsmath,cite,url}
\usepackage{graphicx}
\usepackage{color}
\usepackage{enumitem}

\title{Deep Learning for MIR Tutorial}

\threeauthors
  {Alexander Schindler} { Center for Digital Safety \& Security \\ Austrian Institute of Technology \\ alexander.schindler@ait.ac.at}
  {Thomas Lidy} {Musimap \\ cognitive technologies \\ tom@musimap.com }
  {Sebastian B\"{o}ck} {Faculty of Informatics \\ TU Wien \\ {sebastian.boeck@tuwien.ac.at}}

\begin{document}

\maketitle

\section{Motivation and Format}

Deep Learning has become state of the art in visual computing and continuously emerges into the Music Information Retrieval (MIR) and audio retrieval domain. In order to bring attention to this topic we propose an introductory tutorial on deep learning for MIR. Besides a general introduction to neural networks, the proposed tutorial covers a wide range of MIR relevant deep learning approaches. \textbf{Convolutional Neural Networks} are currently a de-facto standard for deep learning based audio retrieval. \textbf{Recurrent Neural Networks} have proven to be effective in onset detection tasks such as beat or audio-event detection. \textbf{Siamese Networks} have been shown effective in learning audio representations and distance functions specific for music similarity retrieval. We will incorporate both academic and industrial points of view into the tutorial. Accompanying the tutorial, we will create a Github repository for the content presented at the tutorial as well as references to state of the art work and literature for further reading. This repository will remain public after the conference.

\begin{itemize}[noitemsep, topsep=5pt, leftmargin=10pt]
\itemsep0.4em \renewcommand\labelitemi{} \setlength{\itemindent}{-10pt}

\item \textbf{Intended and expected audience:} We propose an introductory tutorial. Thus, we do not target an audience with a particular knowledge in deep learning as well as in audio processing or MIR. We would expect the participants to have experience with the programming language \textit{Python} and preferably with the scientific workflow environment \textit{Jupyter Notebook\footnote{\url{http://jupyter.org/}}}.  Because this is the ISMIR conference, we would expect the audience to have a basic understanding of MIR and the task exemplified in this tutorial, especially music classification, similarity retrieval and onset detection.
%
%
%
%


\item \textbf{Quality}: This tutorial represents an advanced version of \textit{``Deep Learning for Music Classification using Keras''} which will be presented to a general non-audio affine audience at the ML Prague 2018 conference\footnote{\url{https://www.mlprague.com/}}. The proposed tutorial is adapted to address a MIR specific audience. Because we assume a minimal prior knowledge of the MIR problem space from ISMIR attendees and some ideas of how to approach some of the research tasks addressed by the tutorial, we are able to proceed faster on the audio processing parts and to focus more on the application of deep learning on music retrieval tasks. Thus, we are able to include a further segment on onset detection and tempo estimation. Parts of the tutorial as well as the presented Jupyter notebooks and examples are taken from our lectures on MIR, general IR and  machine learning. Consequently, these parts have already been extensively discussed and improved in recent years.

\item \textbf{Software / Hardware Requirements for the participants:} The tutorial makes extensive use of open-source software. It will use both slides and Jupyter\footnote{\url{http://jupyter.org/}} Notebooks for presentation and running example code. The code examples will be given in Python. For the implementation of the neural networks the high-level API provided by the \textit{Keras\footnote{\url{https://keras.io/}}} deep learning library based on a Tensorflow\footnote{\url{https://www.tensorflow.org/}} backend will be used. The accompanying TensorBoard\footnote{\url{https://www.tensorflow.org/programmers_guide/summaries_and_tensorboard}} user-interface will be used to observe a model's course of training. 
The data-sets used for demonstrating the examples will be subsamples of the Million Song Dataset (MSD). To facilitate execution on standard Laptop hardware without GPU support, these datasets will be kept small (approx. 1000 tracks). 

\item \textbf{Any special requirements:} Stable connection to Internet to connect to our GPU servers. Without a reliable Internet connection we will have to scale our examples down so that they can be executed on the CPU of a Laptop computer.

\item \textbf{Originality of topics with respect to previous ISMIR tutorials}: There has not been such a comprehensive MIR deep learning tutorial at the past ISMIR conferences. \textit{ISMIR 2017} featured \textit{T5: Machine-Learning for Symbolic Music Generation} by \textit{Pierre Roy} and \textit{Jean-Pierre Briot} which focused on symbolic music generation and thus differs from the analytical aspects of the proposed tutorial. The introduction to neural networks is similar to part 1 of our outline. We consider this introductory part as highly relevant for the intended target group which is expected to have no or very limited experience with deep learning. Our tutorial also has a strong focus on audio-based music classification and detection approaches using modern neural network techniques (CNNs, RNNs, Siamese Networks) and are covering tasks such as music/speech, music genre and mood classification, onset, tempo and similarity estimation and retrieval.

\item \textbf{Contact Information}: Alexander Schindler. Center for Digital Safety \& Security. AIT Austrian Institute of Technology GmbH. Donau-City-Straße 1, 1220 Vienna, Austria. Mobile: +43 664 8251454. alexander.schindler@ait.ac.at

\end{itemize}

\section{Outline of the tutorial}

\begin{enumerate}

\item \textbf{Deep Learning Basics} (45min):
The tutorial will start with an introduction to deep learning. Based on a brief historical review of neural networks, basic concepts such as activation functions, back-propagation, loss and mini-batch training will be explained. To bring all participants to the same level, basic audio processing techniques will be explained. Using Example 1 we will deepen these concepts and demonstrate how to implement them in Python using the Keras deep learning API.\\
We will advance with introducing different strategies to optimize a neural network, such as Stochastic Gradient Descent and Adam as well as the concept of overfitting and strategies to avoid this - including Dropout, regularization and learning rate decay. Using Example 2 we will introduce basic concepts of audio processing such as time-frequency transformations and perceptual scaling. Further we indicate the relevance of data normalization and introduce basic model evaluation approaches.

\begin{enumerate}

\item \textit{Example 1: Train a simple Fully Connected Neural Network}: The first basic example the understand the basic concepts of how to implement and train a neural network in Keras.

\item \textit{Example 2: Music vs. Speech classification}: Binary classification problem. First basic example using audio data. Introduces audio preprocessing steps, data normalization, different learning optimization strategies and the TensorBoard platform to observe the training progress of neural networks.

\end{enumerate}

\item \textbf{Music Classification} (45min):
This section will introduce Convolutional Neural Networks (CNN) and teach their relevant concepts, such as filter kernels, feature maps and pooling, by means of the traditional MIR tasks of genre and mood classification. We will briefly discuss different CNN architectures, including well known visual computing architectures and the state-of-the-art for MIR. By means of the two examples of this section we will further introduce the concept of transfer learning, by providing a pre-trained neural network for Example 4. 

\begin{enumerate}

\item \textit{Example 3: Music Genre Classification}: Multi-class classification problem. Basic concepts of Convolutional Neural Networks. Training a custom model for a given dataset.

\item \textit{Example 4: Mood Prediction}: Multi-Label classification problem. Basic concepts of transfer learning. Pre-trained model trained on mood labels provided. Fine-tuning model on given dataset. Overview of large-scale industrial music analysis with the example of Musimap\footnote{\url{https://www.musimap.net}}.

\end{enumerate}

\item \textbf{Similarity Estimation and Retrieval} (45min):
This section introduces Siamese Neural Networks as an approach to the task of music similarity retrieval. The participants will sequentially learn how to share CNN layers and how to implement a triplet loss function including appropriate data-preprocessing. 

\begin{enumerate}

\item \textit{Example 5: Siamese Networks using genre labels}: Basic concepts of Siamese Neural Networks. Input triplets are created using genre labels (same genre = similar). Results are presented using listening examples.

\item \textit{Example 6: Siamese Networks using Tag Similarities}: Based on a multi-label ground-truth assignment, a tag-similarity function is applied on the labels to estimate a track-similarity. This similarity is used to create the triplet-inputs for the network. The T-SNE visualization is introduced to demonstrate learned relationships.

\end{enumerate}

\item \textbf{Music Onset Detection and Tempo Estimation} (45min):
Automatic analysis of rhythmic structure in music pieces has been an active research field since the 1970s. It is of prime importance for musicology and tasks such as music transcription, automatic accompaniment, expressive performance analysis, music similarity estimation, and music segmentation.

This segment will provide an overview of the state-of-the-art in computational rhythm description systems, with a special focus on beat and downbeat tracking with recurrent neural networks (RNNs). These systems are usually built, tuned, and evaluated with music of low rhythmic complexity. Based on difficult examples, capabilities of these systems will be analyzed,  their shortcomings investigated, and  challenges and ideas for future research discussed with regard to music of higher complexity.

\begin{enumerate}

\item \textit{Example 7: Audio onset detection in music:} will be presented interactively using \emph{madmom}\footnote{\url{https://github.com/CPJKU/madmom}} \cite{bock2016madmom}, an open source audio signal processing framework written in Python, offering the possibility to get hands-on experience of most algorithms during the tutorial.

\end{enumerate}

\end{enumerate}

\section{Instructor Experience}

\begin{itemize}[noitemsep, topsep=5pt, leftmargin=10pt]
\itemsep0.4em \renewcommand\labelitemi{} \setlength{\itemindent}{-10pt}

\item \textbf{Alexander Schindler:} 

Alexander Schindler is Scientist at the AIT Austrian Institute of Technology and Technical University of Vienna. Since 2010 he is member of the Music Information Retrieval group at the Technical University where he actively participates in research, various international projects and currently finishes his Ph.D on audio-visual analysis of music videos. He actively participates in teaching MIR, machine learning and DataScience where he ambitiously works on transforming the presentation of the lecture-content from static slidedecks to interactive presentations using Jupyter Notebooks and live coding. At the AIT he is responsible for establishing a deep learning group. As task- and project-lead in various audio-related projects he focuses on the application of deep learning approaches to the audio-retrieval domain. Selected relevant articles at international refereed workshops, conferences and journals include \cite{Schindler_2016tist, schindler2016comparing, schindler2016europeana, fazekas2017multi, schindler2017fashion, schindler2017multi}. Alexander is a co-organizer of the monthly Vienna Deep Learning Meetup and chairing the DataScience track of the Semantics 2018 conference.

\item \textbf{Thomas Lidy:}

Thomas Lidy has a long-standing experience in audio analysis and MIR, which he has gathered in more than 12 years as a researcher at TU Wien. His research expertise is on audio feature extraction as well as machine learning for automatic classification of music and clustering and visualization of large music collections. He is the author of 40 articles, book chapters and papers at refereed international workshops and conferences. 
Selected relevant publications include\cite{lidy2016parallel, Lidy_Schindler_MIREX2016, pons_cbmi2016, Lidy2016}. During his time at  TU Wien Thomas was also lecturing undergraduate students in MIR and Deep Learning and supervised students for their thesis.
Thomas actively participated  in the annual MIREX benchmarking initiative, achieving top positions multiple times. More recently, he has a strong focus on Deep Learning for audio, winning 3 benchmark contests with novel neural network based approaches (2x MIREX, DCASE).

Thomas was also a co-organizer of the ECDL 2005, ISMIR 2007 and iPres 2010 conferences as well as Waves Vienna Music Hackday 2015 to 2017. He is the founder and co-organizer of the monthly Vienna Deep Learning Meetup with about 1000 members\footnote{\url{https://www.meetup.com/Vienna-Deep-Learning-Meetup}}.

Thomas had also founded Spectralmind, an innovative music technology company that created both professional music search products and mobile apps for visual music discovery. Currently he is the Head of Machine Learning at Musimap, a large-scale human+AI based music recommender and audio search engine company.

\pagebreak

\item \textbf{Sebastian Boeck:}

Sebastian Boeck received his diploma degree in electrical engineering from the Technical University in Munich in 2010 and his PhD from the Department of Computational Perception at the Johannes Kepler University Linz. Recently he joined the MIR team at  the Technical University of Vienna where he actively participates in research and teaching MIR and DataScience. His main research topic is the analysis of time event series in music signals, with a strong focus on artificial neural networks.
Selected relevant publications include \cite{eyben2010universal,schluter2014improved,bock2012polyphonic,bock2011enhanced,bock2012online,bock2015accurate,bock2016joint}

\end{itemize}

\bibliography{ISMIRtemplate}

\end{document}